\documentstyle[12pt]{article}    
\let\mathrm\rm
\let\mathbf\bf

\def\rect#1#2{{\vcenter{\vbox{\hrule height.3pt
            \hbox{\vrule width.3pt height#2truecm \kern#1truecm
            \vrule width.3pt}
            \hrule height.3pt}}}}

\def\sgn{\mathop{\rm sgn}}
\def\Prob{\mathop{\rm Prob}}

\newtheorem{lemma}{Lemma}[section]

\begin{document}
\begin{center}
{\addtolength{\baselineskip}{\baselineskip} 
\vskip1cm
 {\LARGE 
  Critical droplets in Metastable States of Probabilistic
 Cellular Automata}
\vskip1.0cm
 {\Large Stephen Bigelis$^1$, Emilio N.M.  Cirillo$^2$,
Joel L.  Lebowitz$^{1,3}$ and Eugene R.  Speer$^1$}

}
\vskip0.8cm
  $^1$Department of Mathematics\\ 
  Rutgers University\\
  New Brunswick, NJ 08903\\
\vskip0.3cm
  $^2$Universit\'e Paris Sud\\
  Mathematique, Batiment 425\\
 91405 Orsay Cedex, France\\
\vskip0.3cm
  $^3$Department of Physics\\
  Rutgers University\\
   New Brunswick, NJ 08903 \\
\end{center}

\vskip 1.5 cm 
\begin{abstract} 
We consider the problem of metastability in a
probabilistic cellular automaton (PCA) with a parallel updating rule which
is reversible with respect to a Gibbs measure.  The dynamical rules
contain two parameters $\beta$ and $h$ which resemble, but are not
identical to, the inverse temperature and external magnetic field in a
ferromagnetic Ising model; in particular, the phase diagram of the system
has two stable phases when $\beta$ is large enough and $h$ is zero, and a
unique phase when $h$ is nonzero.  When the system evolves, at small
positive values of $h$, from an initial state with all spins down, the PCA
dynamics give rise to a transition from a metastable to a stable phase when
a droplet of the favored $+$ phase inside the metastable $-$ phase reaches
a critical size.  We give heuristic arguments to estimate the critical size
in the limit of zero ``temperature'' ($\beta\to\infty$), as well as
estimates of the time required for the formation of such a droplet in a
finite system.  Monte Carlo simulations give results in good agreement with
the theoretical predictions.  
\end{abstract}


\newpage

 \section{Introduction}
\label{sec:intr} Metastable states are ubiquitous in systems undergoing
first order phase transitions.  During their lifetime (which can be very
long indeed) these states are practically indistinguishable
from equilibrium states.  Nevertheless, they cannot be described in the
framework of the equilibrium Gibbsian formalism \cite{[PL],[I],[LR]}.
Their analysis in terms of dynamical models has led to a deeper
understanding of metastability by providing detailed descriptions of the
``escape routes'' from metastable to stable states in certain idealized
limiting situations.

\par
 Following earlier work on systems with long range interactions
\cite{[Grif],[PL],[CCO]}, the pathwise approach to metastability was
introduced in \cite{[CGOV]}. It was then used in \cite{[NS]} and
\cite{[BC]} to study rigorously the escape from metastability in the Ising
model with nearest neighbor interactions and a small external magnetic
field, evolving via Metropolis Glauber dynamics in a finite periodic
domain, in the limit of temperature going to zero.  The effect of the
boundary conditions on the exit path from the metastable phase was analyzed
in \cite{[CL]}.  Metastability for spin systems with different interactions
was investigated in \cite{[CO]} and in \cite{[OS]} the problem was
considered in a more general context.
\par

The case of finite temperature, infinite volume and
external magnetic field going to zero was studied in \cite{[S]}; 
this situation, very
interesting from the physical point of view, is 
mathematically much more complicated than the zero temperature limit.
The finite temperature case also has been studied  by means of
Monte Carlo simulations (see e.g. \cite{[B]}) and by transfer-matrix
methods \cite{[GRN]}; a clear discussion 
of these results can be found in \cite{[RTMS]}.
\par

In all the above works (except for that of Penrose-Lebowitz \cite{[PL]}
dealing with deterministic continuum systems) the spin systems evolve
according to a stochastic continuous time or serial dynamics, for which at
most one spin of the system is updated at any time.  In this paper we
investigate metastable behavior in systems with parallel evolution, i.e.,
in which all the spins of the system are updated simultaneously, at integer
times $t=1,2,3,\ldots.$ In particular, we are interested in finding how the
escape time and escape path from the metastable phase are influenced by the
parallel dynamics.  A natural setting for this question is that of
probabilistic cellular automata (PCA), specifically those whose stationary
measures are Gibbs states of Ising models with short range interactions.

PCA were first studied in the Soviet literature of the early seventies
\cite{[St]} and since then have been applied in many different contexts; in
particular, their connections with statistical mechanics were investigated
in \cite{[R],[LMS]}.  In this paper we will consider a PCA for which the
dynamics depends on two parameters, $\beta$ and $h$, and which has the
property that its stationary states are Gibbs measures for a certain
Hamiltonian $H(\beta,h)$.  Here $\beta$ plays the role of an inverse
temperature and $h$ that of an external magnetic field, but the coupling
constants in $H$ depend on $\beta$ in a complicated way; in particular,
$H(\beta,h)\ne\beta\tilde H(h)$.  As in the standard Ising model we have that
when $\beta$ is large enough and $h$ is zero there exist two different
stationary Gibbs measures for the PCA, characterized by nonzero average
magnetizations $\pm m^*$, while for $h \ne 0$ there is a unique stable
phase.  We then pose the usual question of metastability: if at large
$\beta$ the system is prepared in the minus (plus) phase and the magnetic
field is chosen positive (negative) and small, how does the system reach
the stable phase? For definiteness we will always consider the escape from
an initial all minus phase.
 \par

In spin models with continuous time dynamics an important role in this
transition is played by the {\it stable configurations}
\cite{[NS],[BC],[CL],[CO]}, which are fixed points of the evolution in the
limit of zero temperature.  For example, a rectangle of pluses of width
greater than one inside a sea of minuses is a stable configuration for the
nearest neighbor Ising model with a small positive external field.  The
tendency of such a rectangle to grow or to shrink by the repeated addition
or loss of single sites, as a function of its size, yields the behavior of
the exit from the metastable state.  In the PCA discussed below all
configurations are accessible at a single updating.  Nevertheless, we will
argue that, for large $\beta$,  
the only configurations relevant for the description of the exit
from the metastable phase are those in which the plus phase is inside well
separated rectangles in the minus sea.  We then discuss quantitatively the
growth and shrinkage of such droplets and compare our theoretical
prediction with results of Monte Carlo simulations.
 \par

 In Section~\ref{sec:model} we define our model and show that it undergoes
a phase transition at low temperature, and in Section~\ref{sec:special}
describe the specific model on which we will focus in the balance of the
paper.  We discuss our heuristics on the critical behavior of droplets in
Section~\ref{sec:evolution}, and compare theoretical and Monte Carlo
results in Section~\ref{sec:steve}.  Section~\ref{sec:concl} is devoted to
some brief conclusions. 

 \vskip 2 cm
 \section{Description of the General PCA Model}
 \label{sec:model}

Let $\Lambda$ be a $d$-dimensional torus containing $L^d$ lattice sites,
i.e., $\Lambda \subset {\mathbf {Z}^d}$ is a cube containing $L^d$ points and
having periodic boundary conditions.  At each site $x\in\Lambda$ there is
spin variable $\sigma(x)=\pm 1$; the space $\{1,-1\}^\Lambda$ of
configurations is denoted by $\Omega$. 
 \par

To define the dynamics of the model we introduce the discrete time variable
$n=0,1,...$ and denote by $\sigma_n$ the system configuration at time $n$.
All the spins are updated simultaneously and independently at every unit
time; the conditional probability that the spin at site $x$ takes value
$\tau$ at time $n$, given the configuration at time $n-1$, is
 \begin{equation}
   \Prob(\sigma_n(x)=\tau| \sigma_{n-1})
  \equiv p_x(\tau|\sigma_{n-1})
  ={1\over 2} \Bigl[1+\tau \tanh\beta \Bigl(\sum_{y\in\Lambda} K(x-y)
   \sigma_{n-1}(y)+h\Bigr)\Bigr].  \label{eq:rule}
 \end{equation}
 Thus the time evolution is a Markov chain on $\Omega$ with nonzero
transition probabilities
 $P_\Lambda(\eta|\sigma)$ given by
  \begin{equation}
  P_\Lambda(\eta|\sigma)
   =\prod_{x\in\Lambda}p_x\bigl(\eta(x)|\sigma\bigr)\;,\;\;\;\;\;
   \forall\sigma,\eta\in\Omega. \label{eq:markov}
 \end{equation}
 The coupling is of finite range ($K(z)=0$ if $|z|>z_0$, with $z_0<L$ and
typically $z_0\ll L$) and the coupling constants $K(z)$ will be held fixed
throughout our discussion.  The parameters $\beta$ and $h$ play the
role of inverse temperature and external magnetic field respectively, as
discussed above.  Note that for large $|h|$, $\sigma_{n}(x)=\sgn(h)$ with high
probability, while for large positive $\beta$, $\sigma_{n}(x)=\sgn[\sum_y
K(x-y) \sigma_{n-1}(y) + h]$ with high probability. 
 \par

We say that a probability measure $\rho(\sigma)$ on the
configuration space $\Omega$ is {\it stationary} for the PCA if and only if
it remains invariant under the dynamics, i.e., iff $\sum_{\sigma}
P_\Lambda(\eta|\sigma)\rho(\sigma) =
\rho(\eta)$.  By the general theory of Markov processes there
exists, for any $\beta$, $h$, and $\Lambda$, a unique stationary measure
$\nu^{\beta,h}_\Lambda$ for the PCA.  We say that the PCA is {\it reversible}
with respect to a measure $\rho$ iff
 \begin{equation}
 P_\Lambda(\eta|\sigma) \rho(\sigma) =
 P_\Lambda(\sigma|\eta) \rho(\eta)\;,
     \qquad 
\forall\sigma,\eta\in\Omega.    \label{eq:rev}
 \end{equation}
 Summing (\ref{eq:rev}) over $\sigma$ shows that any $\rho$
satisfying (\ref{eq:rev}) is stationary for the PCA; the converse need of
course not be true.  It is however straightforward to check that if $K(z) =
K(-z)$ then the process (\ref{eq:markov}) is reversible with respect to the
measure
 \begin{equation}
  \nu^{\beta,h}_\Lambda(\sigma) = Z^{-1}\prod_{x\in\Lambda}
  e^{\beta h \sigma(x)} \cosh[\beta(\sum_y K(x-y) \sigma(y) + h)]\;,
  \label{eq:meas}
 \end{equation}
 where $Z$ is a normalization constant.  To see this one simply notes 
that $p_x(\eta(x)|\sigma)$ can be written as
 \begin{equation}
  p_x(\eta(x)|\sigma) 
  = {1 \over 2}\;{\exp[\beta h\eta(x)+\beta\sum_y K(x-y)\eta(x)\sigma(y)] 
    \over \cosh \beta[\sum_y K(x-y) \sigma(y) + h]}\;. \label{eq:Gibbs} 
 \end{equation}
 The measure (\ref{eq:meas}) must of course be the unique stationary
measure referred to above.  From (\ref{eq:meas}) it is clear that
$\nu^{\beta,h}_\Lambda(\sigma)$ is a Gibbs measure for a
Hamiltonian $H(\beta,h)$ with (generally many spin) interactions of finite
range, which by our assumptions are independent of $\Lambda$:
 \begin{equation}
  H(\beta,h)(\sigma) = -\beta h\sum_x\sigma(x)
  -\sum_x\log \left[\cosh\beta\left(\sum_y K(x-y)\sigma(y)+h\right)\right]\;.
 \label{eq:Ham}
 \end{equation}
   Hence taking the limit $\Lambda
\nearrow {\mathbf {Z}^d}$ yields a Gibbs measure $\nu^{\beta,h}$ 
for $H$ which is
stationary for the PCA on ${\mathbf {Z}^d}$, defined by the natural extension
of the Markov process (\ref{eq:markov}) to ${\mathbf {Z}^d}$. 

The stationary measures for the infinite volume PCA need of course no
longer be unique.  It is known in general, however, that if one stationary
translation invariant (or periodic) measure is Gibbsian then all such
measures are Gibbsian for the same Hamiltonian \cite{[DaiPra]}.  Hence to
find all translation invariant stationary states of our PCA we need only
investigate translation invariant Gibbs states for $H(\beta,h)$.  Such an
investigation begins with the ground states of the Hamiltonian.  For the
model considered here it is easy to see from (\ref{eq:Ham}) that if
$K(z) \geq 0$ for all $z$, and if the set of $K(z)$ which are nonzero is
not chosen in a very special way, then for $h=0$ there are exactly two
ground states of $H(\beta,h)$, $+\underline1$, in which $\sigma(x)= 1$ for
all $x$, and $-\underline1$, in which $\sigma(x) = -1$ for all $x$, while
for $h \ne 0$ there is only one ground state.  It then follows from the
Pirogov-Sinai theory \cite{[PS]} that for $d \geq 2$ and $\beta$
sufficiently large there will be in general two extremal translation
invariant Gibbs measures for $h=0$ and a unique such measure for $h \ne 0$.
By the argument above, the same conclusion holds for stationary states of
the infinite volume PCA.  We are thus exactly in the same set up as in the
familiar ferromagnetic Ising model.  We remark that although we are not
dealing here with pair interactions, or even exclusively ferromagnetic
interactions, it is easy to see that the measures (\ref{eq:meas}) satisfy
the both the FKG \cite{[FLM]} and GKS \cite{[L]} inequalities.

We may now pose the paradigm question of metastability: if we prepare the
system in the starting configuration $-\underline1$ and take $h$ to be
small and positive, how quickly and in what manner does the PCA reach its
stationary measure?  We want to answer this question in the limit of
$\beta\to\infty$, with $\Lambda$ and $h$ fixed, in which the stationary
state is $+\underline1$, and hence may formulate the first part of the
problem as that of estimating the first hitting time
$\tau_{+1} =\inf\{n\ge 0:\; \sigma_n=+\underline1\}$, in the limit
$\beta\to\infty$, when the system is prepared in $\sigma_0=-\underline1$.
To answer the second part of the question, we need to describe the path
that the system follows to reach $+\underline1$; typically, such a
path will involve the necessity of passing through one of a small
number of critical configurations.
 \par

As in the case of continuous (or serial) dynamics, the first problem is to
understand the behavior of the ``stable'' configurations, that is to estimate
the probability that a stable configuration grows or shrinks.  Rather than
discussing this problem in general we shall now focus on a specific model. 

 \vskip 2 cm \section{A Special Model}
 \label{sec:special} 

 For the rest of this paper we will focus on one special model from among
those specified by (\ref{eq:rule}): the two-dimensional model ($d=2$) with
$K(z)=1$ for $z\in A_0$ and $K(z)=0$ otherwise, where $A_0=\{0,\pm e_1,\pm
e_2\}$ is the set consisting of the origin and its four nearest neighbors. 
Thus the probability distribution of the spin $\sigma_{n}(x)$ is determined by
the spins at time $n-1$ at the five sites in a cross centered at $x$. 
According to (\ref{eq:meas}), the stationary measure $\nu^{\beta,h}_\Lambda$
of this system will then be
 \begin{equation}
 \nu^{\beta,h}_\Lambda(\sigma) =
   Z^{-1}(\beta;\Lambda)\exp[-\sum_{x\in\Lambda} U_x (\sigma;\beta,h)
 + \beta h \sum_{x\in\Lambda} \sigma_x]\;, \label{eq:spmeas}
 \end{equation}
where $U_x(\sigma;\beta,h) = U_0(\tau_{-x}\sigma;\beta,h)$ with $\tau_x$
the shift operator (with periodic boundary conditions on $\Lambda$) and
 \begin{equation}
 U_0(\sigma;\beta,h)
    = - \sum_{A\subset A_0} J_{|A|}(\beta,h) \sigma(A)
     = - \log \cosh (\beta \sum_{y\in A_0} \sigma(y) + \beta h),
   \label{eq:U}
 \end{equation}
with $\sigma(A) = \Pi_{y\in A} \sigma(y)$ for any $A\subset\Lambda$.  The
six coefficients $J_{|A|}(\beta,h)$ are determined by the six values which
the $\sum_{y\in A_0} \sigma(y)$ can take.  For $h=0$ only even values of
$|A|$ occur, and we find
 \begin{equation}
   U_0(\sigma;\beta,0) 
    = -J_0(\beta,0)-J_2(\beta,0) \sum_{\{x,y\}\subset A_0} \sigma_x \sigma_y -
 J_4(\beta,0) \sum_{\{x,y,z,w\}\subset A_0}
     \sigma(x) \sigma(y) \sigma(z) \sigma(w)\;,     \label{eq:U0}
 \end{equation}
 with
 \begin{eqnarray}
 J_0(\beta,0)
  &=& {1 \over 16} \log[(\cosh 5\beta)(\cosh 3\beta)^5(\cosh\beta)^{10}]
      \geq 0\;,
   \label{eq:J0}\\
 J_2(\beta,0)
  &=& {1 \over 16} \log[(\cosh 5\beta)(\cosh 3\beta)/(\cosh\beta)^2] \geq 0\;,
   \label{eq:J2}\\
 J_4(\beta,0) 
  &=& {1 \over 16} \log[(\cosh 5\beta)(\cosh \beta)^2/(\cosh 3\beta)^3] 
   \leq 0\,. \label{eq:J4}
 \end{eqnarray}
 The pair interactions are thus ferromagnetic while the four-spin
interactions are not, so the usual conditions for GKS inequalities are not
satisfied.

\section{The Time Evolution}
\label{sec:evolution} 

We shall first give a heuristic argument showing that the important
configurations for exiting from the metastable state in the PCA are, as for
the usual Glauber dynamics, isolated rectangles of pluses of minimum width
two.  We shall then describe, again on a heuristic level, the growth and
shrinkage of one such rectangle.  We find a critical value $l^*(h)$, for
$h<1$, for the length $l$ of the smaller side of the rectangle such that, in
the limit $\beta \to \infty$, all rectangles with $l < l^*(h)$ will shrink to
zero (except for some special values of $h$, for which the condition is $l <
l^*(h)-1$) while those with $l \geq l^*(h)$ will grow, resulting in an escape
from the metastable state. 

Let us begin by comparing Glauber dynamics---realized via the Metropolis
algorithm, as is usual in questions of metastability---with the PCA
dynamics considered above, focussing on differences which are relevant when
$h$ is small and $\beta$ is very large.  The former, for a spin system with
Hamiltonian $\tilde H=\tilde H(h)$ and inverse temperature $\beta$,
proceeds by spin flips at single sites, with the rate $c(x;\sigma)$ at site
$x$ in configuration $\sigma$ given by
 \begin{equation}
  c(x;\sigma)=\cases{1,& if $\tilde H(\sigma^x)\le\tilde H(\sigma)$,\cr
    \exp\beta[-\tilde H(\sigma^x) + \tilde H(\sigma)],&
    if $\tilde H(\sigma^x)>\tilde H(\sigma)$,\cr}
 \end{equation}
 where $\sigma^x$ is the configuration obtained from $\sigma$ by flipping
the spin at site $x$:
 \begin{equation}
  \sigma^x(y) = \cases{-\sigma(y),& for $y=x$,\cr
       \sigma(y),& for $y \ne x$.\cr}  
 \end{equation}
 Since $c(x;\sigma)$ depends only on
$\beta[\tilde H(\sigma^x) - \tilde H(\sigma)]$ and is independent of
$\beta$ if $\tilde H(\sigma^x)\le\tilde H(\sigma)$, the dynamical landscape
is determined entirely by the function $(\beta H)({\bf \sigma})$, and the
{\it stable configurations}, i.e., those invariant under the dynamics in
the limit $\beta \to \infty$, are the local minima of $\tilde H$.

In contrast, the PCA dynamics permits transitions from one configuration to
any other in a single updating; we will see, however, that this distinction
will play only a minor role in the analysis of metastability.  Recall that
the probability of a transition from $\sigma$ to another configuration $\eta$
in one time step is given by the product of the probabilities of spin flips
at sites where $\sigma$ and $\eta$ differ with the probabilities of non-flips
at sites where they agree.  Probabilities of all possible single site flips
are shown in Figure~\ref{fig:meccanismi}; it is clear that, at large $\beta$,
certain flips are almost sure to take place, while all others have
exponentially small probability.  Thus from an arbitrary initial condition we
expect a very rapid evolution to a stable configuration, with further change
taking place on an exponentially slow time scale, and involving primarily
flips of single spins---in fact, on six different exponential time scales,
well separated for $h<1$ and very large $\beta$, corresponding to the six
slow spin flip processes of Figure~\ref{fig:meccanismi}.  The parallelism of
the dynamics is of relevance during the first, rapid, phase, but the analysis
of metastability involves the slow, essentially serial, second
phase---although in certain cases we must consider the effect of a small
number of unlikely single-site events occurring simultaneously. 

A second difference is that, although in the PCA dynamics transitions which
lower the energy are generally favored over those which do not, the
probability of a transition from $\sigma$ to $\eta$ is not specified
entirely by the energy difference
$H(\eta; \beta, h) - H({\sigma}; \beta, h)$.  In particular, for the
specific model introduced in Section~\ref{sec:special}, the single-step
probability of flipping a spin which agrees with two of its nearest
neighbors is exponentially small, even when such a flip is (energetically)
favored by the magnetic field, and hence there are pairs of configurations
${\bf \sigma}$ and ${\bf \eta}$ which differ at a single site but are such
that the probability of jumping between them (in either direction) goes to
zero as $\beta \to \infty$.  This is illustrated in
 Figure~\ref{fig:meccanismi}.  Consequently there are many more stable
configurations for the PCA than for the Glauber dynamics.  In fact it is
easy to see that any configuration in which the value of the spin at every
site agrees with that of at least two of its neighboring sites is stable.

Despite the large number of stable configurations, however, it is
rectangular droplets which are important for exit from the metastable
state, due to the effect of the most rapid of the ``slow'' single flip
processes of Figure~\ref{fig:meccanismi}.  We formalize this as follows.
Let $\widehat\Omega$ be the set of configurations in which every plus spin
agrees with at least two of its nearest neighbors, and define
$T:\widehat\Omega\to\widehat\Omega$ so that $T\sigma$ is the configuration
obtained from $\sigma$ by flipping all the minus spins with at least two
pluses among their nearest neighbors.  For any $\sigma\in\widehat\Omega$
the sequence of configurations $T^k\sigma$, $k=0,1,\ldots,$ is nondecreasing,
in the sense that $(T^{k+1}\sigma)(x)\ge (T^k\sigma)(x)$ for all $x$,
and hence must reach a fixed point $\sigma^*$, in which the set of plus
spins forms well separated rectangles (or bands around the torus) inside
the sea of minuses.  Moreover, if we take $\sigma$ as the initial condition
$\sigma_0$ of the PCA dynamics, and let $E_\sigma$ be the event that
for some $n$, $\sigma_n=\sigma^*$ and  the sequence $(\sigma_m)$ is
increasing for $0\le m\le n$, then it is clear from
 Figure~\ref{fig:meccanismi} that
 \begin{equation}
  \lim_{\beta\to\infty}\Prob E_\sigma = 1,
 \end{equation}
 and that the time to reach $\sigma^*$ is typically of order
$\exp{2\beta(1-h)}$.  Thus the path for escape from metastability must pass
through configurations in which all the pluses are inside well separated
rectangles.  

We do not attempt to discuss the most general situation but instead
consider the fate of a single rectangle; we expect that, as for Glauber
dynamics \cite{[NS]}, this is the key element in an analysis of
metastability.  Let us consider, then, a configuration $\eta$ for which all
spins inside a rectangle of sides $l$ and $m$ are up and all other spins
are down, and suppose for definiteness that $l\le m$. We will say that such
an $l\times m$ droplet is {\it supercritical} if, starting from $\eta$, the
system will reach the configuration $+\underline1$ before it reaches
$-\underline1$, with probability which approaches 1 in the limit
$\beta\to\infty$; the droplet is {\it subcritical} if the reverse is true.
We will argue heuristically that if $l<2/h$ then the droplet is subcritical
and if $l>2/h$ the droplet is supercritical, while if $2/h$ is an integer
and $l=2/h$ then the droplet may either shrink or grow.  The {\it critical
length} $l^*_h$ is the smallest integer such that the droplet is
supercritical if $l\ge l^*_h$; thus $l^*_h=\lfloor2/h\rfloor+1$.

A very rough estimate of $l^*_h$ may be based on energy considerations,
with the assumption that if the system starts from a rectangular droplet
then the next droplet reached will be one with lower energy. For example,
if $\eta$ contains an $l\times l$ droplet and $e(l)=H(\eta)$, one may
approximate $e(l)$ at very large $\beta$ by writing
$\log\cosh\beta x\approx\beta |x|$ in (\ref{eq:U}).  In this approximation,
$e(l)$ is a parabola whose maximum is achieved at $l=2/h$, supporting the
result $l^*_h=\lfloor2/h\rfloor+1$ described above.

 For a correct calculation of the critical length we must analyze in detail
the mechanisms of growth and shrinkage of a rectangular droplet; these are
in general similar to those for Glauber dynamics \cite{[NS]}, although the
details are different.  We will comment below on the possibilities of
making  the following heuristic discussion rigorous.

Consider first growth.  From Figure~\ref{fig:meccanismi} it is clear that a
single plus protuberance on one of the four sides of the rectangular is not
stable; growth proceeds through the formation of a double protuberance,
which then grows ``quickly'' (i.e., on the time scale $\exp{2\beta(1-h)}$)
to complete the additional side.  The parallel dynamics permits the double
protuberance to form in one time step, as shown in
 Figure~\ref{fig:growth-shrink}(a); the typical time for this process is
\begin{equation} 
\tau_{\mathrm {double}}\sim {\mathrm e}^{4\beta (3-h)}.
\label{eq:growth-double}
\end{equation}
 Alternatively, the protuberance can
grow in two consecutive time steps, as shown in
 Figure~\ref{fig:growth-shrink}(b); the parallel character of the dynamics
enters here as well, since after formation of a single protuberance at the
first step there must occur, at the second step, both the persistence of
this protuberance, with probability $\exp(-2\beta (1-h))$, and the flip of
a minus spin adjacent to the protuberance, also with probability
$\exp(-2\beta (1-h))$ (see Figure~\ref{fig:meccanismi}); the typical time
for this growth process is thus
\begin{equation} 
\tau_{\mathrm {single}}\sim 
\underbrace{{\mathrm e}^{2\beta (3-h)}}_{\mathrm {first\; step}}
\times
\underbrace{{\mathrm e}^{4\beta (1-h)}}_{\mathrm {second\; step}}
\label{eq:growth-single}
\end{equation}
 Clearly $\tau_{\mathrm {single}}\ll \tau_{\mathrm {double}}$ for $\beta$
large and hence the most efficient growth mechanism is the two-step one.

Again from Figure~\ref{fig:meccanismi} it is clear that the most efficient
shrinking mechanism is the usual corner erosion, shown in
 Figure~\ref{fig:growth-shrink}(c); the shrinking is performed via a
sequence of stable configurations.  We estimate the time needed for the
loss of one of the shorter sides of the rectangle, which requires the
erosion of $l-1$ sites (after which the remaining single protuberance
vanishes rapidly).  When $\beta$ is large, such a process will typically
occur without backtracking.  The rate at which the entire process occurs is
thus estimated as the rate for one erosion, $\exp^{-2\beta (1+h)}$, times
the probability that $l-2$ further erosions occur within the lifetime
$\exp^{2\beta (1-h)}$ of a stable configuration, which is of order
$\Bigl[\exp^{-2\beta (1+h)}\exp^{2\beta (1-h)}\Bigr]^{l-2}$.  Thus the
shrinking time is estimated as
\begin{equation} \tau_{\mathrm {shrink}}\sim {\mathrm e}^{2\beta (1+h)}
\times 
\left[
\frac{{\mathrm e}^{2\beta (1+h)}}{{\mathrm e}^{2\beta (1-h)}}\right]
^{l-2}.
\label{eq:shrink}
\end{equation}

\begin{figure}
\begin{center}
\begin{picture}(200,100)
\put(0,0){\line(1,0){60}}
\put(0,0){\line(0,1){70}}
\put(0,70){\line(1,0){50}}
\put(60,0){\line(0,1){40}}
\put(50,40){\line(1,0){10}}
\put(50,40){\line(0,1){30}}
\put(75,50){$\uparrow\; {\mathrm e}^{2\beta (1-h)}$}
\put(75,25){$\downarrow\; {\mathrm e}^{2\beta (1+h)}$}
\end{picture}
\end{center}
\vskip 2.5 cm 
\caption{Approximate description of the behavior of the edge of a droplet.}
\label{fig:birth}
\end{figure}
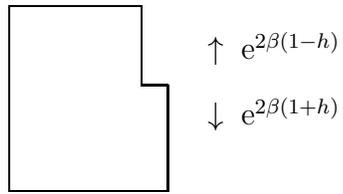

The estimate of the shrinking time $\tau_{\mathrm {shrink}}$ 
can be supported by considering a random walk  which
models what happens on one edge of the droplet. Consider
a Markov chain $X_t$, $t=0,1,2,...$, taking values in the nonnegative
integers and with transition probabilities
\begin{displaymath}
P(k,l)=\left\{
\begin{array}{cl}
\frac{1}{2}{\mathrm e}^{-\beta c} & {\mathrm {if}}\; l=k+1,\\
&\\
\frac{1}{2}{\mathrm e}^{-\beta b} & {\mathrm {if}}\; l=k-1,\\
&\\
1-\frac{1}{2}{\mathrm e}^{-\beta c}
-\frac{1}{2}{\mathrm e}^{-\beta b} & {\mathrm {if}}\; l=k,\\
&\\
0 & {\mathrm {otherwise},}\\
\end{array}
\right.
\end{displaymath}
and
\begin{displaymath}
P(0,1)=\frac{1}{2}{\mathrm e}^{-\beta c}\;,\;\;\;\;\;\;
P(0,0)=1-\frac{1}{2}{\mathrm e}^{-\beta c},
\end{displaymath}
where $k\ge 1$, $c>b>0$, and $\beta>0$.  This chain, with $b=2(1-h)$ and
$c=2(1+h)$, is an approximate description of the behavior of the edge of a
droplet (see Figure~\ref{fig:birth}), if one thinks of $X_t$ as
representing the number of minus spins on this edge at time $t$. In order
to estimate $\tau_{\mathrm {shrink}}$ one should calculate the typical time
to see $l-1$ minus spins on the edge, starting from zero minus spins.  Such
an estimate, which agrees with (\ref{eq:shrink}), is provided by the
following lemma, whose proof is parallel to that of Lemma~1 of \cite{[NS]}.
\begin{lemma}
\label{lemma:birth} For $k\ge1$, define the hitting time $\tau^0_k$ for
the Markov chain $X_t$ with $X_0=0$ by
 \begin{equation}
\tau^0_k\stackrel{\mathrm{def}}{=}\{t\ge 1:\; X_t=k\}.
\label{eq:hitt_birth}
 \end{equation}
Then for any  $\varepsilon>0$,
\begin{equation}
P\left(
{\mathrm e}^{\beta ck-\beta b(k-1) -\beta\varepsilon}
<\tau^0_k<
{\mathrm e}^{\beta ck-\beta b(k-1) +\beta\varepsilon}
\right)
\stackrel{\beta\to\infty}{\longrightarrow}  1\;.
\end{equation}
\end{lemma}

To complete the derivation of the critical length for rectangular droplets
we compare (\ref{eq:growth-single}) and (\ref{eq:shrink}): growth occurs
with probability one in the zero temperature limit, that is, $l\ge l^*_h$,
if $\lim_{\beta\to\infty}\tau_{\mathrm {single}}/\tau_{\mathrm {shrink}}=0$.
This again leads to
 \begin{equation}
l^*_h=\left\lfloor\frac{2}{h}\right\rfloor +1 \label{eq:critical_infty}.
 \end{equation}

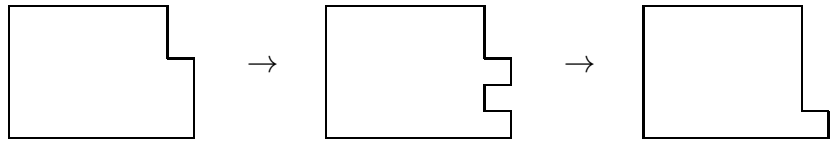
\begin{figure}
\begin{center}
\begin{picture}(360,50)
\put(0,0){\line(1,0){70}}
\put(0,0){\line(0,1){50}}
\put(0,50){\line(1,0){60}}
\put(70,0){\line(0,1){30}}
\put(60,30){\line(0,1){20}}
\put(60,30){\line(1,0){10}}
\put(120,0){\line(1,0){70}}
\put(120,0){\line(0,1){50}}
\put(120,50){\line(1,0){60}}
\put(190,0){\line(0,1){10}}
\put(180,10){\line(1,0){10}}
\put(180,10){\line(0,1){10}}
\put(180,20){\line(1,0){10}}
\put(190,20){\line(0,1){10}}
\put(180,30){\line(1,0){10}}
\put(180,30){\line(0,1){20}}
\put(240,0){\line(1,0){70}}
\put(240,0){\line(0,1){50}}
\put(240,50){\line(1,0){60}}
\put(310,0){\line(0,1){10}}
\put(300,10){\line(0,1){40}}
\put(300,10){\line(1,0){10}}
\put(90,25){$\rightarrow$}
\put(210,25){$\rightarrow$}

\end{picture}
\end{center}

\vskip 2.5 cm
\caption{One mechanism occurring on time scale faster than
$\tau_{\mathrm {single}}$ and hence relevant for complete treatment of
$\tau_{\mathrm {shrink}}$.}
\label{fig:two-step-shrink}
 \end{figure}

We believe that the above argument could be made rigorous along the lines
of the corresponding arguments in Section~2 of \cite{[NS]}.  The main
complicating factor appears to be that, because the growth time
$\tau_{\mathrm {single}}$ is so large, processes beyond simple corner
erosion must be accounted for when evaluating the shrinking time.  For
example, several corners may disappear in one step; more complicated
processes, such as the two-step shrinkage by two sites shown in
 Figure~\ref{fig:two-step-shrink}, are also relevant.  These modifications
appear to be technical only and should not affect the estimate
(\ref{eq:shrink}) of $\tau_{\mathrm {shrink}}$, but we have not carried out
a complete analysis.

\vskip 2 cm
\section{Monte Carlo results}
\label{sec:steve}

 The critical length $l^*_h$ introduced in the previous section characterizes
the behavior of the system in the limit $\beta\to\infty$.  In this section we
define a critical length $l^*_{\beta,h}$ at finite $\beta$ and describe the
results of Monte Carlo simulations evaluating $l^*_{\beta,h}$ numerically for
several values of $\beta$ and $h$.  We find that when $\beta$ is large enough
the resulting estimates of $l^*_{\beta,h}$ are close the theoretical estimate
of $l^*_h$ given in the previous section. 
 \par

Let $p_{\beta,h}(l)$ denote the probability  that a square
droplet of side $l$ grows and covers the whole lattice, that is, that in
evolving from this initial configuration the system reaches the state
$+\underline1$ before the state $-\underline1$.  Clearly $p_{\beta,h}(l)$
is a nondecreasing function of $l$ with 
$p_{\beta,h}(0)=0$
and $p_{\beta,h}(L)=1$, so that the differences
\begin{equation}
d_{\beta,h}(l)=p_{\beta,h}(l)
-p_{\beta,h}(l-1)
\label{eq:deriv}
\end{equation}
 form a normalized probability distribution.  In the limit $\beta\to\infty$
(assuming for simplicity that $2/h$ is not an integer), $p_{\beta,h}$ reduces
to a step function,
 \begin{equation}
p_{\infty,h}(l)=\left\{
\begin{array}{cl}
1,&{\mathrm{if}}\; l\ge l^*_h,\\
0,&{\mathrm{if}}\; l< l^*_h,\label{pinfty}
\end{array}
\right.
\label{eq:prob_inf}
 \end{equation}
 and $d_{\beta,h}(l)$ to a unit mass on the critical length $l^*_h$.  At 
finite temperature, then, we define the 
critical length to be the mean of the distribution $d_{\beta,h}$:
 \begin{equation}
l^*_{\beta,h}=\sum_l l d_{\beta,h}(l).
\label{eq:mean}
 \end{equation}
 From (\ref{pinfty}) it follows that $l^*_{\infty,h}=l^*_h$.  We note
that this approach in the numerical estimate of the critical length
is different from that used in \cite{[CL]}.

In the above discussion we have suppressed the dependence of $p_{\beta,h}$,
$d_{\beta,h}$, and $l^*_{\beta,h}$ on the lattice size $L$, since for large
$\beta$ and $L$ we expect $l^*_{\beta,h}$ as defined by (\ref{eq:mean}) to be
essentially independent of $L$. 

We have carried out numerical experiments to estimate the function
$p_{\beta,h}(l)$, and hence $l^*_{\beta,h}$, for $\beta=0.9,1.1,1.3$ and
$h=0.05,0.1,0.2$; we varied $l$ over the range of values in which
$p_{\beta,h}(l)$ changes rapidly and made $N=100$ runs for each value of $l$. 
 For each run, we first prepared our system in a starting configuration
characterized by a single square droplet of plus spins of size $l$, placed in
a lattice whose size $L$ was chosen large enough to avoid boundary effects. 
We then followed the evolution of the system and decided, by means of lower
and upper cutoffs on the total system magnetization, whether the droplet
would ultimately grow or shrink.  We also introduced a cutoff on the total
length of each run, chosen as a function of $\beta$ so that for most runs the
fate of the droplet was determined before the cutoff was reached; in the case
$\beta= 1.3$, the highest value of $\beta$ we have considered, this cutoff
was $200,000$ iterations.  Letting $G_{\beta,h}(l)$ denote the number of
times that the droplet grew and $S_{\beta,h}(l)$the number of times that it
shrank, and assuming that the fraction of the remaining
$N-G_{\beta,h}-S_{\beta,h}$ cases (in which we did not determine the
behavior) in which the droplet would have grown if we had waited long enough
is the same as for the cases in which the behavior was determined, we are led
to the estimate
 \begin{equation}
p_{\beta,h}(l)=\frac{G_{\beta,h}(l)}{G_{\beta,h}(l)+S_{\beta,h}(l)}.
\label{eq:estim_growth}
 \end{equation}

 From the estimated values of $p_{\beta,h}(l)$ we computed the finite
temperature critical length, via (\ref{eq:mean}), and  the standard
deviation of the distribution $d_{\beta,h}$; the values are recorded in
Table \ref{tab:tabella}.  The results are in a very good agreement with our
theoretical prediction: when the temperature is lowered, the numerical
measure of the critical length tends to the zero temperature theoretical
prediction $l^*_h=\lfloor2/h\rfloor+1$.  We did not consider higher values
of $\beta$ because too long runs would have been needed, but the values we
have considered seem to be sufficient to see the zero temperature limit
behavior.  Note that the standard deviation 
of the distribution $d_{\beta,h}(l)$
decreases when $\beta$ is increased.  This good behavior is clearest in the
case of small external magnetic field; presumably, higher values of $\beta$
should be considered at higher $h$ to approach the limiting behavior.

We observed that the typical time for growth of the initial square depended
strongly on $\beta$ and $h$, but not on $l$; while the typical shrinking time
increase sensibly when $l$ is increased.  This is qualitatively in agreement
with theoretical estimates (\ref{eq:growth-single}) and (\ref{eq:shrink}). 

\vskip 2 cm
\section{Conclusions}
\label{sec:concl}

In this paper we have studied the problem of metastable states in
probabilistic cellular automata, viewing the latter as the simplest instance
of models evolving under parallel dynamics.  All detailed work has been
focussed on a particular case: a two dimensional model on the square lattice
in which the probability of a spin flip at site $x$ depends only on the total
magnetization of the set of five spins in a cross centered at $x$ (see
(\ref{eq:rule})). 

We conclude that the general pattern of analysis which has been used for
similar models evolving under Glauber dynamics applies here as well, since
events in which the system makes a one-step transition to a configuration
significantly different from the current one can be neglected in the
low-temperature limit.  In particular, we argue that the path of escape from
metastability passes through a critical rectangular droplet.  On the other
hand, the parallel nature of the dynamics does influence the details of the
analysis of the escape time and, in particular, adds enough complications
to make a rigorous analysis more difficult than in the Glauber case.

 For the model in question we have shown, through heuristic arguments and
Monte Carlo simulations, that the critical length of a rectangular droplet is
$l^*_h=\lfloor2/h\rfloor+1$.  Our theoretical prediction is valid only in the limit of
zero temperature, but our simulations confirm estimates close to the
theoretical ones even at finite temperature.

It is natural to ask whether escape from the metastable state is
facilitated or hindered by the use of parallel (as opposed to serial)
dynamics.  It is not clear that this question has a universal answer, but
as a preliminary approach we may ask what would happen in the model of this
paper if a serial evolution rule were adopted, so that at each time step
one spin is chosen at random, with uniform probability, and then updated
with probability given by~(\ref{eq:rule}).

As remarked in Section~\ref{sec:evolution}, we used the parallel character of
the dynamics only in the estimate of $\tau_{{\mathrm single}}$, so that to
estimate of the critical length in the serial case one should compare the
shrinking time (\ref{eq:shrink}) with a new growing time $\tau_{\mathrm
{growth}}\sim \exp [2\beta (3-h)+2\beta(1-h)]$, obtained by noting that in
the second step of the double protuberance growth the persistence probability
of the single protuberance need be taken into account.  Comparison of these
two times shows that the critical length in the serial case is given, for $h$
very small, by $\lfloor3/2h\rfloor+1$.  Thus, for these models, 
the parallel rule leads to a larger critical droplet and a slower
exit from the metastable phase.

\vskip 2 cm
 \par
\noindent
{\Large\bf Acknowledgments}
\vskip 0.3 cm
 \par
\noindent 
One of the authors (E.C.) wishes to express his thanks  
to the Mathematics Department of Rutgers
University for its very kind hospitality and to Enzo Olivieri
for useful discussions. E.C. also thanks Istituto Nazionale di Fisica 
Nucleare - Sezione di Bari and Dipartimento di Fisica dell'Universit\`a 
degli Studi di Bari for their financial support. Work at Rutgers
was supported by NSF Grant DMR 95 - 23266. JLL would also like to thank
DIMACS and its supporting agencies the NSF under contract STC--91--19999
and the NJ Commission on Science and Technology.  We thank Alex Mazel for
very useful discussions.

\newpage
\par
\noindent
\begin{table}
\begin{center}
\begin{tabular}{cccccc}      \hline\hline
   &      &      &$\beta$&       &    \\
   &      &$0.9$&$1.1$&$1.3$&$\infty$\\ \hline
   &$0.05$&$38.53\pm 1.60$&$40.16\pm 1.50$&$40.58\pm 1.35$&$41$\\ \cline{3-6}
$h$&$0.1$ &$19.76\pm 1.13$&$20.29\pm 1.20$&$20.36\pm 0.92$&$21$\\ \cline{3-6}
   &$0.2$ &$ 9.96\pm 0.20$&$ 9.96\pm 0.20$&$ 9.98\pm 0.14$&$11$\\ \hline\hline
\end{tabular}
\end{center}
\vskip 2 cm

\caption{Estimates of the critical length $l^*_{\beta,h}$,
with their standard deviations, obtained from Monte Carlo
simulations via the procedure described in Section \ref{sec:steve}. For
$\beta=\infty$ we give $l^*_h$ as obtained from (\ref{eq:critical_infty}).}
\label{tab:tabella}
\end{table}
$\phantom .$

 \par
\noindent
\vskip 2 cm
\begin{figure}
\begin{center}
\begin{tabular}{ccccccccccc} 
\begin{picture}(24,24)(0,+16)
\put(12,0){$+$}
\put(0,12){$+$}
\put(12,12){$+$}
\put(12,24){$+$}
\put(24,12){$+$}
\end{picture}
&& ${1\over 1+e^{2\beta(5+h)}}\sim e^{-2\beta(5+h)}$ &&&&&&  
\begin{picture}(12,12)(0,+16)
\put(12,0){$-$}
\put(0,12){$-$}
\put(12,12){$-$}
\put(12,24){$-$}
\put(24,12){$-$}
\end{picture}
&& ${1\over 1+e^{2\beta(5-h)}}\sim e^{-2\beta(5-h)}$ 
\\
&&&&&&\\
&&&&&&\\
\begin{picture}(24,24)(0,+16)
\put(12,0){$-$}
\put(0,12){$+$}
\put(12,12){$+$}
\put(12,24){$+$}
\put(24,12){$+$}
\end{picture}
&& ${1\over 1+e^{2\beta(3+h)}}\sim e^{-2\beta(3+h)}$ &&&&&&
\begin{picture}(12,12)(0,+16)
\put(12,0){$+$}
\put(0,12){$-$}
\put(12,12){$-$}
\put(12,24){$-$}
\put(24,12){$-$}
\end{picture}
&& ${1\over 1+e^{2\beta(3-h)}}\sim e^{-2\beta(3-h)}$
\\
&&&&&&\\
&&&&&&\\
\begin{picture}(24,24)(0,+16)
\put(12,0){$-$}
\put(0,12){$+$}
\put(12,12){$+$}
\put(12,24){$+$}
\put(24,12){$-$}
\end{picture}
&& ${1\over 1+e^{2\beta(1+h)}}\sim e^{-2\beta(1+h)}$ &&&&&&
\begin{picture}(12,12)(0,+16)
\put(12,0){$+$}
\put(0,12){$-$}
\put(12,12){$-$}
\put(12,24){$-$}
\put(24,12){$+$}
\end{picture}
&& ${1\over 1+e^{2\beta(1-h)}}\sim e^{-2\beta(1-h)}$
\\
&&&&&&\\
&&&&&&\\
\begin{picture}(24,24)(0,+16)
\put(12,0){$-$}
\put(0,12){$+$}
\put(12,12){$+$}
\put(12,24){$-$}
\put(24,12){$-$}
\end{picture}
&& ${1\over 1+e^{-2\beta(1-h)}}\sim 1 - e^{-2\beta(1-h)}$ &&&&&&
\begin{picture}(12,12)(0,+16)
\put(12,0){$+$}
\put(0,12){$-$}
\put(12,12){$-$}
\put(12,24){$+$}
\put(24,12){$+$}
\end{picture}
&& ${1\over 1+e^{-2\beta(1+h)}}\sim 1 - e^{-2\beta(1+h)}$
\\
&&&&&\\
&&&&&\\
\begin{picture}(24,24)(0,+16)
\put(12,0){$-$}
\put(0,12){$-$}
\put(12,12){$+$}
\put(12,24){$-$}
\put(24,12){$-$}
\end{picture}
&& ${1\over 1+e^{-2\beta(3-h)}}\sim 1 - e^{-2\beta(3-h)}$ &&&&&&
\begin{picture}(12,12)(0,+16)
\put(12,0){$+$}
\put(0,12){$+$}
\put(12,12){$-$}
\put(12,24){$+$}
\put(24,12){$+$}
\end{picture}
&& ${1\over 1+e^{-2\beta(3+h)}}\sim 1 - e^{-2\beta(3+h)}$
\\
\end{tabular}
\end{center}
\vskip 4 cm
\caption{Probabilities for the flip of 
the central spin for all 
possible configurations in the $5$-spin neighborhood.}
\label{fig:meccanismi}
\end{figure}
$\phantom .$

 \par
\noindent
\vskip 2 cm
\begin{figure}
\begin{center}
\begin{picture}(360,50)
\put(0,0){\line(1,0){70}}
\put(0,0){\line(0,1){50}}
\put(0,50){\line(1,0){70}}
\put(70,0){\line(0,1){50}}
\put(120,0){\line(1,0){70}}
\put(120,0){\line(0,1){50}}
\put(120,50){\line(1,0){70}}
\put(190,0){\line(0,1){20}}
\put(190,40){\line(0,1){10}}
\put(200,20){\line(0,1){20}}
\put(190,20){\line(1,0){10}}
\put(190,40){\line(1,0){10}}
\put(240,0){\line(1,0){80}}
\put(240,0){\line(0,1){50}}
\put(240,50){\line(1,0){80}}
\put(320,0){\line(0,1){50}}
\put(90,25){$\rightarrow$}
\put(215,25){$\rightarrow$}
\put(-10,20){$l$}
\put(30,-10){$m$}
\put(150,-10){$m$}
\put(265,-10){$m+1$}
\end{picture}
\vskip 0.8 cm
(a)
\vskip0.8cm
\begin{picture}(440,50)
\put(0,0){\line(1,0){70}}
\put(0,0){\line(0,1){50}}
\put(0,50){\line(1,0){70}}
\put(70,0){\line(0,1){50}}
\put(120,0){\line(1,0){70}}
\put(120,0){\line(0,1){50}}
\put(120,50){\line(1,0){70}}
\put(190,0){\line(0,1){20}}
\put(190,30){\line(0,1){20}}
\put(200,20){\line(0,1){10}}
\put(190,20){\line(1,0){10}}
\put(190,30){\line(1,0){10}}
\put(240,0){\line(1,0){70}}
\put(240,0){\line(0,1){50}}
\put(240,50){\line(1,0){70}}
\put(310,0){\line(0,1){20}}
\put(310,40){\line(0,1){10}}
\put(320,20){\line(0,1){20}}
\put(310,20){\line(1,0){10}}
\put(310,40){\line(1,0){10}}
\put(360,0){\line(1,0){80}}
\put(360,0){\line(0,1){50}}
\put(360,50){\line(1,0){80}}
\put(440,0){\line(0,1){50}}
\put(90,25){$\rightarrow$}
\put(215,25){$\rightarrow$}
\put(335,25){$\rightarrow$}
\put(-10,20){$l$}
\put(30,-10){$m$}
\put(150,-10){$m$}
\put(270,-10){$m$}
\put(390,-10){$m+1$}
\end{picture}
\vskip 0.8 cm
(b)
\vskip0.8cm
\begin{picture}(425,50)
\put(0,0){\line(1,0){70}}
\put(0,0){\line(0,1){50}}
\put(0,50){\line(1,0){70}}
\put(70,0){\line(0,1){50}}
\put(100,0){\line(1,0){70}}
\put(100,0){\line(0,1){50}}
\put(100,50){\line(1,0){60}}
\put(170,0){\line(0,1){40}}
\put(160,40){\line(1,0){10}}
\put(160,40){\line(0,1){10}}
\put(200,0){\line(1,0){60}}
\put(200,0){\line(0,1){50}}
\put(200,50){\line(1,0){60}}
\put(260,0){\line(0,1){10}}
\put(260,10){\line(1,0){10}}
\put(260,40){\line(1,0){10}}
\put(260,40){\line(0,1){10}}
\put(270,10){\line(0,1){30}}
\put(355,0){\line(1,0){60}}
\put(355,0){\line(0,1){50}}
\put(355,50){\line(1,0){60}}
\put(415,0){\line(0,1){10}}
\put(415,10){\line(1,0){10}}
\put(425,10){\line(0,1){10}}
\put(415,20){\line(1,0){10}}
\put(415,20){\line(0,1){30}}
\put(80,25){$\rightarrow$}
\put(180,25){$\rightarrow$}
\put(280,25){$\rightarrow$}
\put(300,25){$\cdots\cdots$}
\put(335,25){$\rightarrow$}
\put(-10,20){$l$}
\put(30,-10){$m$}
\put(130,-10){$m$}
\put(230,-10){$m$}
\put(385,-10){$m$}
\end{picture}
\vskip 0.8 cm
(c)
\vskip0.8cm
\end{center}

\vskip 2.5 cm
\caption{Growth and shrinking mechanisms: 
(a)~double protuberance growth mechanism, 
(b)~single protuberance growth mechanism, 
(c)~corner erosion.}
\label{fig:growth-shrink}
 \end{figure}
 $\phantom .$

\newpage
 \par
\noindent
\vskip 2 cm
 $\phantom .$

\newpage
 \par

\noindent
\vskip 2 cm
 $\phantom .$

\end{document}